\documentstyle[12pt,lathuile]{article}
\begin{document}
\title{ 
PRELIMINARY RESULTS AND PERSPECTIVES IN THE ARCHEOPS EXPERIMENT
}
\author{
Jean-Christophe Hamilton        \\
{\em Physique Corpuculaire et Cosmologie, Coll{\`e}ge de France, Paris, France} \\
On behalf of the {\sc Archeops} Collaboration
}
\maketitle
\baselineskip=14.5pt
\begin{abstract}
  Observations of the Cosmic Microwave Background (CMB) temperature
  fluctuations are a powerful tool for testing theories of the early
  Universe and for measuring cosmological parameters.  We present
  basics of CMB physics, review some of the most recent results and
  discuss their implications for cosmology. The Archeops balloon-borne
  experiment is designed to map the CMB with an angular resolution of
  10 arcminutes and a precision of 30 $\mu\mathrm{K}$ per pixel. This
  will allow the measurement of the CMB fluctuation power spectrum
  from large to small angular scales. We describe preliminary results
  from the test flight which took place in July 1999 and present
  perspectives for upcoming scientific flights (January 2001).
\end{abstract}
\baselineskip=17pt
\newpage
\section{Basics of CMB physics}
Within the frame of Big-Bang theory, the Universe started about ten
billion years ago in an extremely hot and dense phase.  The dynamics
of the Universe are governed by the Einstein equation that links the
evolution of the space-time metric to the matter content of the
Universe. Our Universe appears highly homogeneous and isotropic; this
simplifies the metric to the so-called Robertson-Walker
metric\cite{robertson,walker}, which depends on the scale factor of
the Universe and its curvature which can be positive, negative or
zero. The Einstein equations are simplified to the Friedman equations,
which show that the Universe is expanding. The expansion can be
eternal or not, depending on the value of the total matter density
$\Omega_m$ and the cosmological constant $\Omega_\Lambda$.

Just after the Big-Bang, the temperature of the Universe was so hot
that all the matter in the Universe was ionized. As the mean free path
of photons in ionized matter is small, the photons were in thermal
equilibrium with the baryons. The Universe was therefore optically
thick. As it expanded, the temperature of the Universe cooled down and
more and more electrons started to be captured by baryons. Because of
the large photon to baryon ratio ($\simeq 10^9$), the photons belonging
to the high energy tail of the Planck distribution kept the matter
ionized even after the temperature of the Universe dropped below
$13.6$ eV ($160000$ K). But when the temperature cooled down to $0.3$
eV ($3000$ K), most of the matter in the Universe became neutral. This
moment is known as recombination. As the Universe was no longer
ionized, the photon mean free path became larger than the horizon,
and the photons have not interacted with matter since. This is why this
moment is also known as the matter-radiation decoupling.

Those photons that last scattered at this epoch have now cooled down
to a temperature of $2.7$ K (they were redshifted by a factor $\simeq
1100$). They have a pure blackbody spectrum as they were at thermal
equilibrium before decoupling. We see them as a homogeneous and
isotropic radiation called the Cosmic Microwave Background (hereafter
CMB).  Their present density is about $412$ photons.cm$^{-3}$.

The CMB was discovered accidentally in 1965 by Penzias and
Wilson\cite{penziaswilson} and was immediately interpreted as a relic
radiation of the Big-Bang by Dicke and its collaborators\cite{dicke}.
Such a radiation had been predicted before by Gamow\cite{gamow} and by
Alpher and Herman\cite{alpherherman} in 1948. This discovery was a major
argument in favor of the Big-Bang theory. In 1992, the American
satellite COBE discovered the first anisotropies in the temperature of
the CMB with an amplitude of about 30 $\mu$K\cite{cobe}.  COBE also
measured its spectrum with high precision\cite{firas1,firas2} proving
its pure blackbody nature.  Such temperature fluctuations were
expected, as they correspond to density fluctuations in the early
Universe.  We know that density fluctuations existed in the early
Universe because a perfectly homogeneous Universe cannot lead to
structure formation, galaxies, etc.

\begin{figure}[t]
 \vspace{10.5cm}
\includegraphics{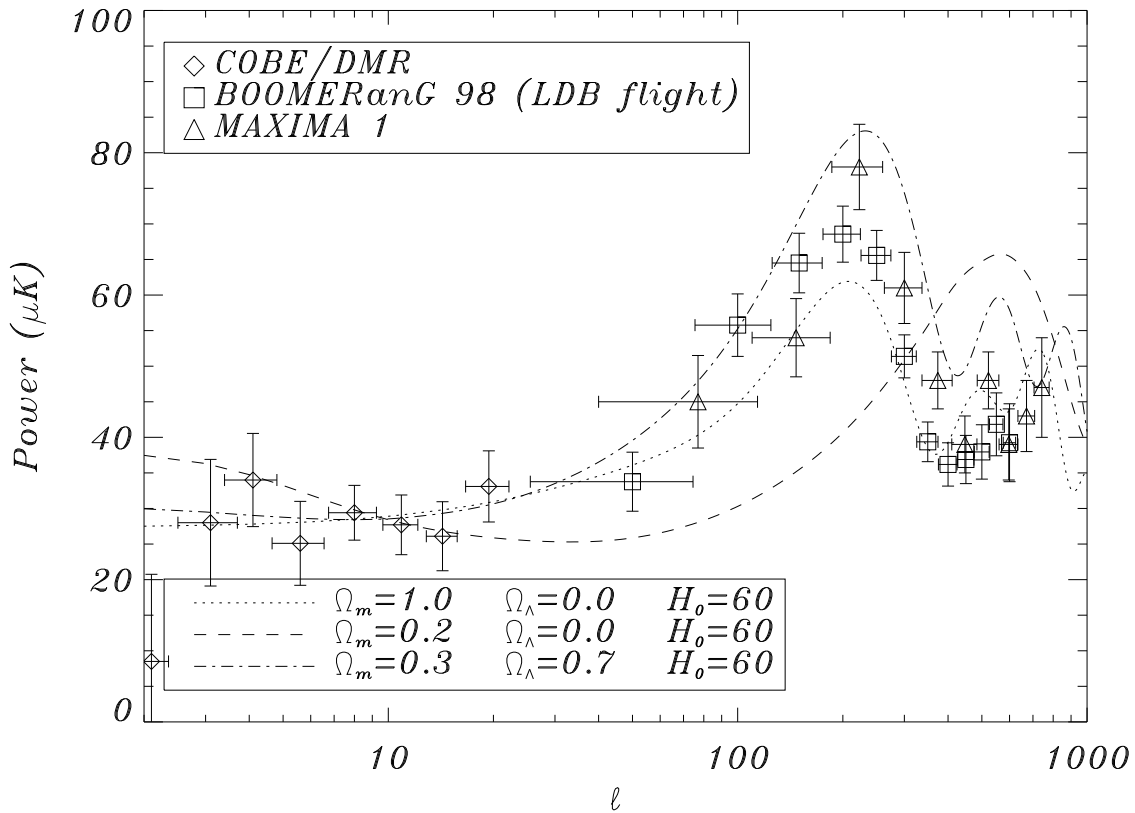}
 \caption{\it Theoretical predictions for the angular power spectrum of the 
   anisotropies of the CMB (obtained using CMBFast\cite{cmbfast} for
   inflationary like models) : for a critical (zero curvature)
   Universe with no cosmological constant (dotted line), for an open
   Universe with no cosmological constant (dashed line) and for a
   critical Universe with a dominating cosmological constant
   (dot-dashed line). The data points are from COBE\cite{cobe},
   from BOOMERanG\cite{boomerangomega} and from MAXIMA\cite{maxima}).
    \label{clfig} }
\end{figure}

The CMB temperature fluctuations are commonly described using a
spherical harmonic expansion (which is the analogue of a Fourier
transform on the sphere):
\begin{equation}
\frac{\delta T}{T}(\theta,\phi)=\sum_{\ell=0}^\infty \sum_{m=-\ell}^\ell
a_{\ell m} Y_{\ell m}(\theta,\phi)
\end{equation}
where $\ell$ is the multipole index and is the analogue of the Fourier
mode $k$ and $\ell$ is inversely proportional to angular scale (1 degree
roughly corresponds to $\ell=200$). The angular power spectrum of the
temperature fluctuations of the CMB is defined as :
\begin{equation}
C_\ell=\frac{1}{2\ell+1}\sum_{m=-\ell}^\ell | a_{\ell m}| ^2.
\end{equation}
The statistical distribution and the angular power spectrum of the
temperature fluctuations of the CMB are predicted to be very different
depending on whether these fluctuations arose from inflation or from
topological defects.  This gives us a tool with which to obtain
information about the early Universe. In the case of inflation, one
can derive the power spectrum as a function of the cosmological
parameters\footnote{Freely available numerical codes, such as
  CMBFast\cite{cmbfast}, have been developped for this purpose.}.
Fig.~\ref{clfig} represents theoretical predictions for the angular
power spectrum of the temperature fluctuations of the CMB for three
different cosmological models (all of them suppose inflationary-like
origins for the density fluctuations). The data point overplotted are
from the most recent experiments.
The first thing that appears when looking at the theoretical
predictions for power spectra is that there are two different parts.
The first part is at low values of $\ell$ (large angular scales) and
exhibits no particular features. It is known as the Sachs-Wolf
plateau.  These large angular scales correspond to regions larger than
the horizon at the time of decoupling. There has therefore not been
any physical process coupling these regions since the end of
inflation: the power spectrum is flat in this region. The second part
of the power spectrum is the high $\ell$ region corresponding to
angular scales smaller than the horizon at the time of decoupling
(around $\ell=200$). These regions have been in causal contact before
decoupling and we expect physical processes to have occurred. The
physical processes involved are acoustic oscillations in the
photon-baryon fluid before decoupling: the baryons tend to collapse
because of their gravitational interaction, but the photon radiation
pressure does not allow any structure to collapse. There are then a
series of acoustic oscillations in the fluid (collapse leading to
overdensities, expansion leading to underdensities and so on)
involving larger and larger regions of the Universe as the horizon is
increasing with time. This process ends when the photons decouple from
the baryons, which are now allowed to collapse and form structures.
Depending on the status (overdensity, underdensity or average density)
of regions of the Universe of a given size, there will be peaks or
valleys in the power spectrum.  This effect is mitigated to some
degree by the Doppler effect arising from the speed of matter in
collapsing or uncollapsing regions at the moment of decoupling.  The
series of peaks we see in the angular power spectrum of the CMB are
therefore called {\em acoustic peaks} and are just the result of these
acoustic oscillations that happened in the photon-baryon fluid before
decoupling.

The position and amplitude of the acoustic peaks (and particularly of
the first one, corresponding to the size of the sound horizon at the
time of decoupling) as seen from here and now depend upon the values
of the cosmological parameters. Basically, we are looking at regions
of a given physical size located at redshift $z\simeq 1100$. The
apparent angular size of these regions (where the acoustic peaks
appear in the power spectrum) depends on the curvature of the Universe
between $z=1100$ and us which involves the cosmological parameters.
For instance, a low density Universe would have a negative curvature
so that the same physical size would be seen at lower angular size
from a cosmological distance than in a critical density Universe. The
first acoustic peak would therefore be shifted to larger values of
$\ell$. This can be seen in Fig. \ref{clfig} where the first acoustic
peak for the dashed curve (low density Universe) occur at
$\ell\simeq500$ whereas it occurs at $\ell\simeq200$ for the critical
$\Omega=1$ models (dotted and dash-dotted curves).

We see that the experimental data points from COBE\cite{cobe} confirm
the theoretical predictions of a flat shape of the power spectrum at
low $\ell$ while the most recent experimental results on smaller
angular scales (BOOMERanG\cite{boomerangomega} and
MAXIMA\cite{maxima}, two balloon-borne experiments) confirm the
existence of a peak in the power spectrum at $\ell\simeq 200$,
strongly favoring a flat ($\Omega=1$) Universe\cite{boomerang}.

What can also be seen in Fig. \ref{clfig} is that COBE concentrated
its measurements on large angular scales ($\ell \le 20$) whereas the
high resolution new experiments concentrate on small angular scales.
This can be easily explained as COBE was a satellite that covered the
whole celestial sphere with a very large beam (7 degrees FWHM), it
could therefore only constrain the low $\ell$ part of the power
spectrum as larger values of $\ell$ correspond to angular scales
smaller than the beam. On the other hand, high resolution experiments
like BOOMERanG or MAXIMA have a very small beam (less than $15$
arcminutes FWHM) allowing the measurement of large values of $\ell$
but they concentrate on a small part of the sky in order to obtain a
large signal to noise ratio and a large redundancy. The {\sc Archeops}
experiment is intended to measure the CMB power spectrum from small to
large values of $\ell$ allowing to link COBE data to the first
acoustic peak and beyond with a single experiment.
\section{The {\sc Archeops} experiment}
{\sc Archeops}\cite{archweb} is a balloon-borne experiment involving
teams from France, Italy, UK and USA. The sky temperature is measured
with cold bolometers and our scanning strategy allows us to cover a
large fraction of the sky (25\%). The instantaneous sensitivity and
angular resolution of the {\sc Archeops} experiment is similar the
that of the other balloon-borne experiments such as BOOMERanG and
MAXIMA.  The optical concept, read-out electronics and cryogenic
system are very similar to what will be used for the bolometer
instrument (High Frequency Instrument) on the {\sc Planck}
satellite\footnote{{\sc Planck} is an ESA satellite designed to make
  high precision measurements of the CMB fluctuations. The satellite
  will be launched in 2007.}. {\sc Archeops} therefore both takes
advantage of {\sc Planck} developments and gives the opportunity to
validate hardware, software and methods for {\sc Planck} on real data.

\begin{figure}[t]
 \vspace{8.5cm}
\includegraphics{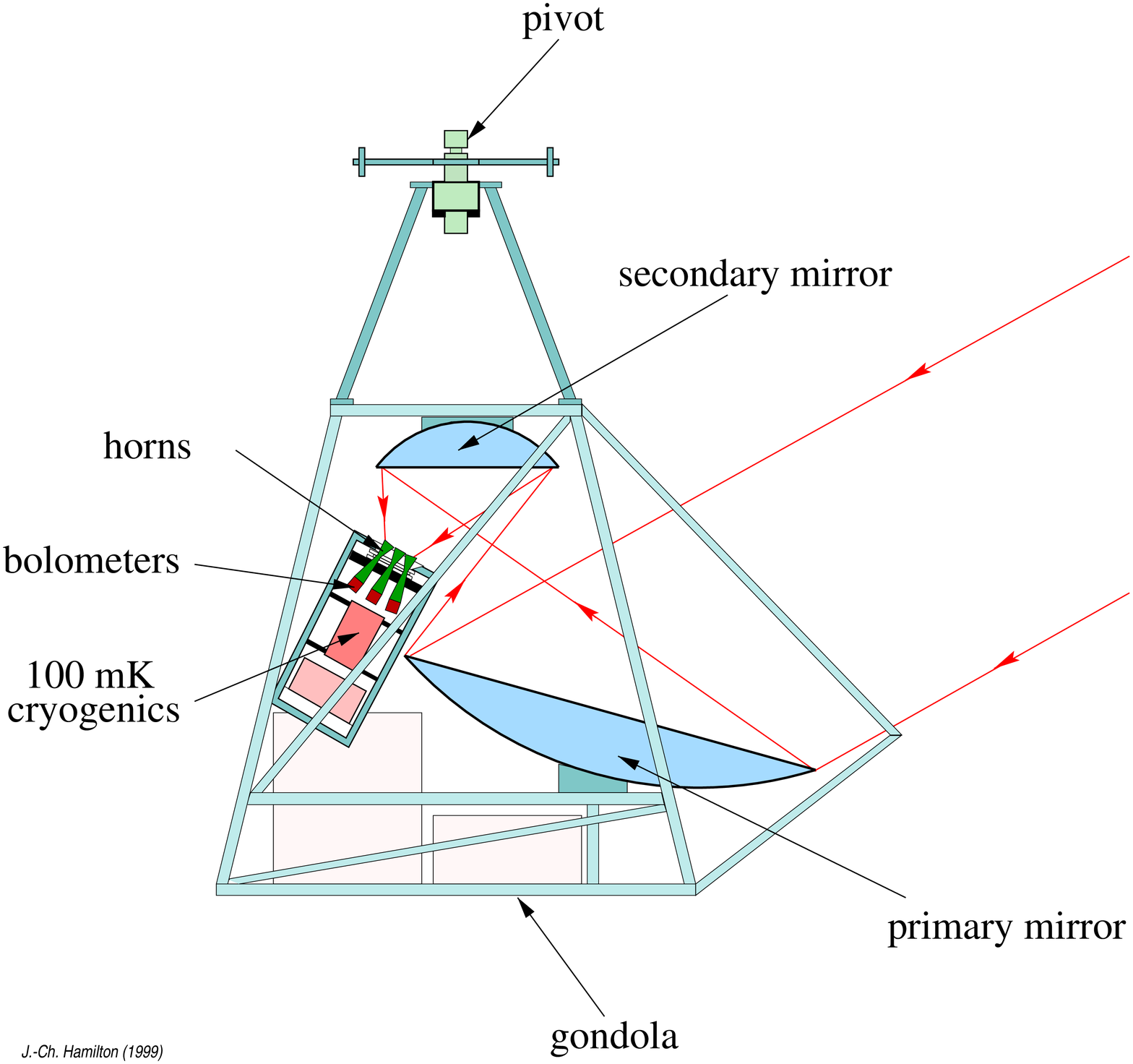}
 \caption{\it The {\sc Archeops} experiment.
    \label{gondola} }
\end{figure}

The {\sc Archeops} gondola hangs about 100 meters under a
stratospheric balloon. The diameter of the balloon, at the cruise
altitude of 40 km, is 120 meters.  The gondola and the instruments are
shown in Fig~\ref{gondola}.
The photons are first collected by the primary mirror (1 meter
effective diameter) and then directed by the secondary mirror towards
the horns, which are designed to filter the correct frequency bands
and to focus the photons on the bolometers. During the flight, the
gondola rotates at 3 revolutions per minute\footnote{The period will be 2 revolutions per minute for the scientific flights.} so that the beam draws
large circles on the sky at a constant elevation (about $41^\circ$).
As the Earth rotates, the part of the sky seen by the detectors slowly
shifts in such a way that a growing fraction of the sky is swepted.

The detectors are spider-web bolometers\footnote{Bolometers with a
  spider-web shape designed to have a small effective surface in order
  to be less sensitive to the cosmic rays.}\cite{bock_bolos} at 4
different frequencies (143, 217, 353 and 545 GHz). The bolometers are
cooled down to 100 mK using a dilution cryostat.

\begin{figure}[t]
 \vspace{3.5cm}
\includegraphics{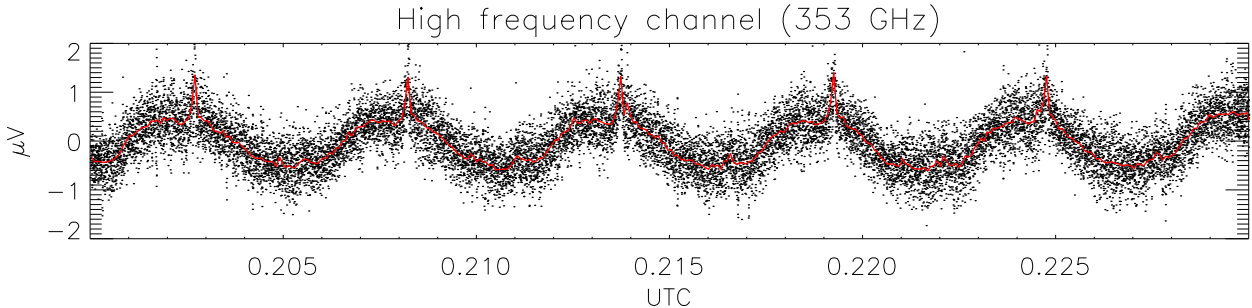}
 \caption{\it Sample time-ordered data for the high frequency channel (353 GHz).
    \label{timeline} }
\end{figure}

Archeops had a successful test flight in July 1999 between Trapani
(Sicily) and Granja de Torre Hermosa (Spain). The focal plane
contained 6 bolometers (one channel failed due to wiring problems) at
3 different frequencies: 143, 217 and 353 GHz. We obtained 4 hours of
good quality night-time data covering 17\% of the sky. The analysis of
these data is in progress. A sample part of the time-ordered data we
obtained is shown in Fig.~\ref{timeline}. The voltage output of the
bolometer (proportional to the temperature) is shown as a function of
Universal Time for a few rotations of the gondola. The first thing we
see is a large oscillation which is scan synchronous. Each rotation of
the gondola corresponds to a period in the timeline. The bright and
narrow peak seen at each rotation is due to Galactic plane crossing.
This large amplitude is mainly a parasitic signal due to reflections
from the balloon and is lower at lower frequencies. This parasitic
signal will have to be removed before any CMB extraction.

\begin{figure}[t]
 \vspace{9cm}
\includegraphics{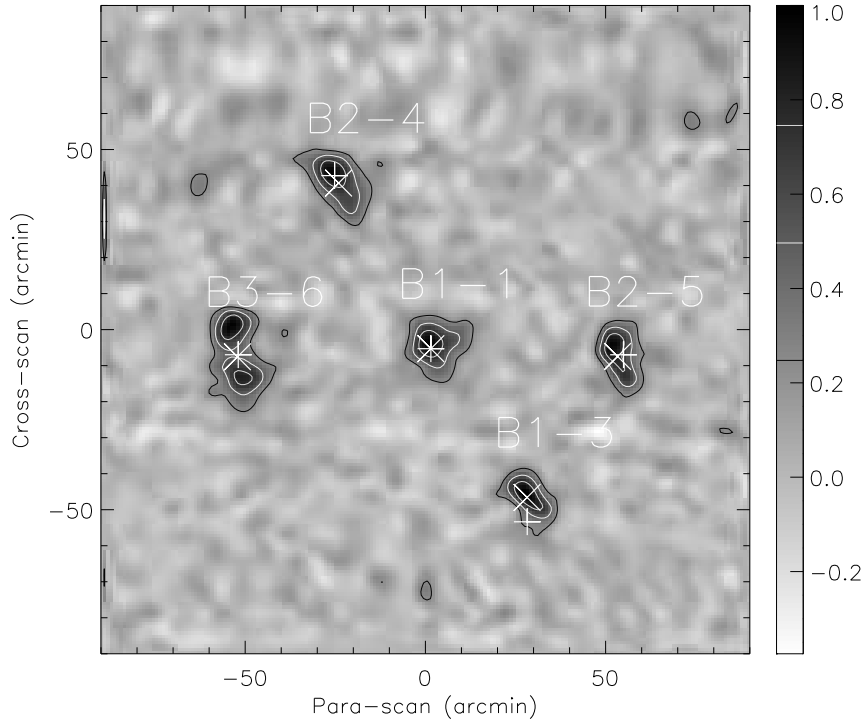}
 \caption{\it Beam shapes measured on Jupiter represented at their respective 
   position within the focal plane. Bolometer 1-1 (middle) and 1-3
   (bottom) are at 143 GHz, bolometers 2-4 (top) and 2-5 (right) are
   at 217 GHz and bolometer 3-6 (left) is at 353 GHz. Bolometer 2-5
   exhibits a much larger amount of high frequency noise than the
   others (due to wiring problems) and will not be used for analysis.
   The + signs indicate theoretical pre-flight positions and the
   $\times$ signs indicate the fitted Gaussian center for each
   bolometer.
    \label{focal} }
\end{figure}

The response of our instrument to a point source (shape of the beam)
has been measured inflight on Jupiter. The angular size of Jupiter
(less than 1 arcminute) is much smaller than our resolution so that it
is effectively a point source for our instrument. The image we obtain
from Jupiter is then the convolution of its true profile on the sky (a
Dirac peak) by the instrumental response. It is therefore a map of our
effective beams.  These maps are shown in Fig.~\ref{focal},
represented at their respective position in the focal plane (also
measured with Jupiter).
The beam widths all turn out to be smaller than 13 arcminutes, except
for bolometer 3-6 which is very elongated in the cross-scan direction
and exhibits a double peak. This is not surprising as for this channel
the feed horn was not corrugated. Moreover, Jupiter's brightness
temperature is known with an accuracy of about 5\%, allowing us to
perform a point source calibration.

\begin{figure}[ht]
 \vspace{5cm}
\includegraphics{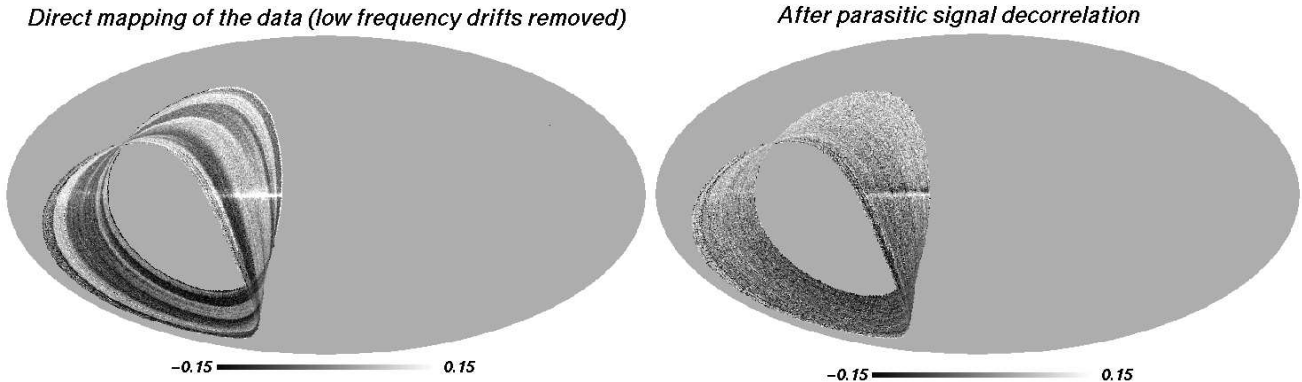}
 \caption{\it Direct reprojection of the data (143 GHz channel, low 
   frequency drifts removed) on the celestial sphere (left panel).
   The right panel shows (with the same color scale) the data after
   parasitic signal decorrelation (using the 353 GHz channel as a
   tracer of the parasitic signal).  The dipole is now clearly visible
   and can be used for calibration. The pixellisation used for these maps
   is HEALPix\cite{healpix}.
    \label{parasitic} }
\end{figure}

In Fig.~\ref{parasitic}, we show the reprojection of our data on the
celestial sphere (after low frequency drift removal using Fourier
filtering on the timelines). The effect of the parasitic signal is
clearly visible. To remove this parasitic signal, we developed a
method based on the fact that it is correlated between different
frequency bands (it is higher at higher frequencies). The 353 GHz
channel is then used as a tracer of the parasitic signal.  The right
panel of Fig.~\ref{parasitic} shows the map obtained after
decorrelation of this effect. The large remaining gradient
is that part of our data correlated to the dipole (due to the
Doppler effect of our motion with respect to the CMB frame). The
dipole is an extended source with very well known temperature and can
therefore be used for calibration.

\begin{figure}[!t]
 \vspace{3.4cm}
\includegraphics{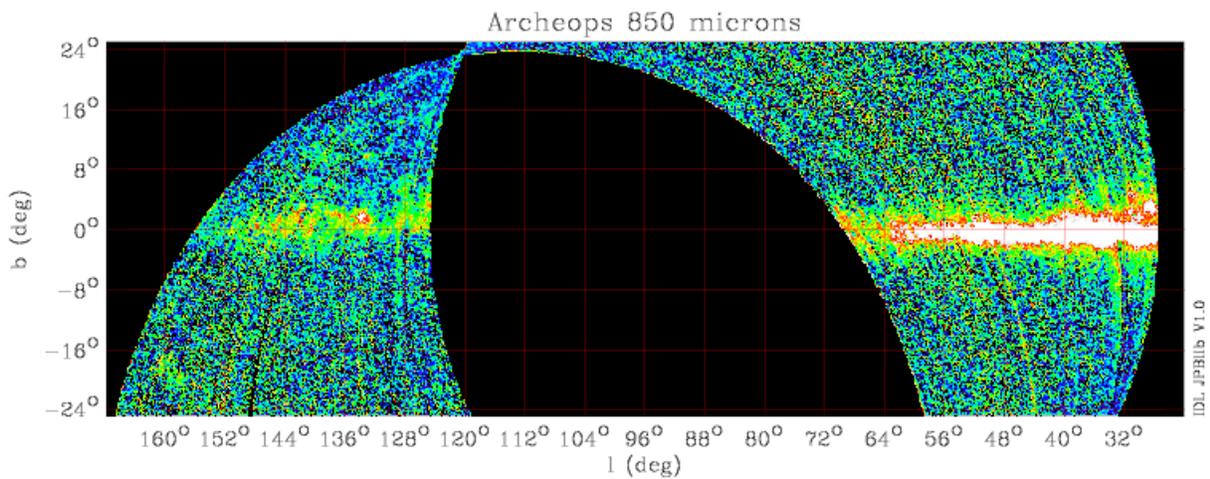}
 \caption{\it Preliminary map of the galactic plane in the {\sc Archeops}
   353 GHz channel (850 microns wavelength) This is the highest
   resolution map at this frequency. Bright sources can already be
   seen at high galactic latitude.
    \label{galaxy} }
\end{figure}

Besides the estimation of cosmological parameters, CMB experiments
produce maps of the sub-millimeter sky that are of high interest for
Galactic astronomy and for cosmology. We obtained with the Archeops flight
test the highest resolution maps of the galactic plane at 353 GHz. A
preliminary map of the Galactic plane is shown in Fig.~\ref{galaxy}
and is now under study.  High latitude sources can be already seen in
this preliminary map.

\section{Conclusions and perspectives}
{\sc Archeops} had a successful test flight last summer and produced 4
hours of high quality data (with 4 bolometers) on 17\% of the sky. The
angular resolution achieved during the test flight was better than 13
arcminutes and we expect it to be 10 arcminutes for future flights.
The experiment will fly this winter from Kiruna, Sweden yielding at
least 24 hours of night-time data with 24 bolometers in 4 frequency
bands and a sky coverage of 25\%. This future flight will provide us
much more data than the test flight allowing a better estimation of
the systematic effects (parasitic signals, temperature drifts ...).

With a very large ratio between sky coverage and angular resolution,
{\sc Archeops} will provide high quality maps of the CMB sky
($\simeq400 000$ independent pixels) allowing a precise measurement of
the CMB angular power spectrum from large angular scales ($\ell\simeq
10$) to small angular scales ($\ell\simeq 800$). Measuring the power
spectrum over such a large range will allow us to measure the
cosmological parameters with high precision and to start constraining
the physics of the early Universe.

\newpage%

\end{document}